\documentclass[aps, prl, twocolumn,superscriptaddress,nofootinbib,10pt]{revtex4-1}
\usepackage{graphicx}
\usepackage{dcolumn}
\usepackage{amssymb,amsmath,amsthm,mathrsfs}
\usepackage{epstopdf}
\usepackage{lipsum}
\usepackage{hyperref}
\usepackage{cleveref}
\usepackage{empheq}
\usepackage{color}
\usepackage[normalem]{ulem} 
\usepackage{physics}
\usepackage[normalem]{ulem}
\usepackage[ruled,vlined]{algorithm2e}

\newcommand{\be}{\begin{equation}}
\newcommand{\ee}{\end{equation}}
 \newcommand{\bea}{\begin{eqnarray}}
\newcommand{\eea}{\end{eqnarray}}

\begin{document}

\vspace*{0.5cm}

\title{{``Stairway to Heaven"} -- Spectroscopy of Particle Couplings with Gravitational Waves}

\newcommand{\addressIFIC}{Instituto de F\'isica Corpuscular (IFIC), Universitat de Val\`{e}ncia-CSIC, E-46980, Valencia, Spain}
\newcommand{\addressXX}{ Center for Nuclear Theory, Department of Physics and Astronomy,  Stony Brook University, New York 11794, USA.}
\newcommand{\addressICL}{Theoretical Physics, Blackett Laboratory, Imperial College, London, SW7 2AZ, United Kingdom}

\author{Daniel G. Figueroa} \affiliation{\addressIFIC}
\author{Adrien Florio} \affiliation{\addressXX}
\author{Nicol\'as Loayza} \affiliation{\addressIFIC}
\author{Mauro Pieroni\,} \affiliation{\addressICL}

\date{\today}

\begin{abstract}
We discuss the possibility to measure particle couplings with stochastic gravitational wave backgrounds (SGWBs). 
Under certain circumstances a sequence of peaks of different amplitude and frequency -- a {\it stairway} --, emerges in a SGWB spectrum, with each peak probing a different coupling. The detection of such signature opens the possibility to reconstruct couplings (spectroscopy) of particle species involved in high energy phenomena generating SGWBs. Stairway-like signatures may arise in causally produced backgrounds in the early Universe, e.g.~from preheating or first order phase transitions. As a proof of principle we study  
a preheating scenario with an inflaton $\phi$ coupled to multiple {\it daughter} fields $\lbrace \chi_j \rbrace$ 
with different coupling strengths.
As a clear stairway signature is imprinted in the SGWB spectrum, we reconstruct the relevant couplings with various detectors.
\end{abstract}

\keywords{cosmology, early Universe, inflation, gravitational waves, stochastic gravitational wave backgrounds}

\maketitle

\noindent {\bf Introduction}. Gravitational wave (GW) astronomy has emerged as an exciting new field initiated by the detection of
GWs by the LIGO/Virgo network, yielding outstanding results on the relativistic dynamics of compact binaries, and enabling stringent tests of gravity, see e.g.~\cite{LIGOScientific:2016aoc,LIGOScientific:2016vlm,LIGOScientific:2018mvr,Venumadhav:2019lyq,LIGOScientific:2020ibl,LIGOScientific:2017vwq,LIGOScientific:2017zic}. This new observational probe offers as well unprecedented opportunities for breakthroughs in high-energy physics and early universe cosmology. The key point to note is that early universe dynamics operate at energies unreachable by any terrestrial means, sourcing stochastic GW backgrounds (SGWBs) that carry information about the phenomena that created them. Cosmological SGWBs of diverse origin are in fact expected to permeate the Universe, see~\cite{Caprini:2018mtu,Maggiore:2018sht} for reviews. 

As supported by cosmic microwave background (CMB) observations~\cite{Akrami:2018odb}, we assume an early phase of accelerated expansion, {\it inflation}, as the framework to explain the initial conditions of the universe~\cite{Guth:1980zm, Linde:1981mu, Albrecht:1982wi,Brout:1977ix,Starobinsky:1980te,Kazanas:1980tx,Sato:1980yn,Mukhanov:1981xt, Guth:1982ec,Starobinsky:1982ee, Hawking:1982cz}. SGWBs arise naturally due to quantum vacuum fluctuations in vanilla inflation models~\cite{Grishchuk:1974ny,Starobinsky:1979ty, Rubakov:1982df,Fabbri:1983us}, as well as from complicated dynamics of axion-like species present during inflation, see e.g.~\cite{Anber:2006xt,Sorbo:2011rz,Pajer:2013fsa,Adshead:2013qp,Adshead:2013nka,Maleknejad:2016qjz,Dimastrogiovanni:2016fuu,Namba:2015gja,Ferreira:2015omg,Peloso:2016gqs,Domcke:2016bkh,Caldwell:2017chz,Guzzetti:2016mkm,Bartolo:2016ami,DAmico:2021zdd,DAmico:2021vka}. Inflation is followed by a \textit{(p)reheating} stage, converting the energy available into particle species that eventually dominate the energy budget~\cite{Allahverdi:2010xz,Amin:2014eta,Lozanov:2019jxc,Allahverdi:2020bys}. SGWBs are expected from particle production during preheating~\cite{Easther:2006gt,GarciaBellido:2007dg,GarciaBellido:2007af,Dufaux:2007pt,Dufaux:2008dn,Dufaux:2010cf,Bethke:2013aba,Bethke:2013vca,Figueroa:2017vfa,Adshead:2018doq,Adshead:2019lbr,Adshead:2019igv}, and from various post-inflationary phenomena, like kination-domination~\cite{Giovannini:1998bp,Giovannini:1999bh,Boyle:2007zx,Figueroa:2018twl,Figueroa:2019paj,Gouttenoire:2021jhk}, oscillon dynamics~\cite{Zhou:2013tsa,Antusch:2016con,Antusch:2017vga,Liu:2017hua,Amin:2018xfe}, strong first order phase transitions~\cite{Kamionkowski:1993fg,Caprini:2007xq,Huber:2008hg,Hindmarsh:2013xza,Hindmarsh:2015qta,Caprini:2015zlo,Hindmarsh:2017gnf,Cutting:2018tjt,Cutting:2018tjt,Cutting:2019zws,Pol:2019yex,Cutting:2020nla}, or cosmic defects~\cite{Vachaspati:1984gt,Sakellariadou:1990ne,Damour:2000wa,Damour:2001bk,Damour:2004kw,Figueroa:2012kw,Hiramatsu:2013qaa,Blanco-Pillado:2017oxo,Auclair:2019wcv,Figueroa:2020lvo,Caprini:2019egz,Gorghetto:2021fsn,Chang:2021afa}. 

A detection program including a large variety of experiments, and covering a wide range of frequencies, is slowly emerging. These experiments range from increasingly precise CMB and pulsar timing array observations~\cite{Abazajian:2016yjj,Arzoumanian:2020vkk}, to present and proposed GW direct detection experiments~\cite{Sesana:2019vho,LISA:2017pwj,Ruan:2020smc,Baker:2019pnp,Kawamura:2011zz,Somiya:2011np,Mei:2020lrl,Kuns:2019upi,Reitze:2019iox,Punturo:2010zz,LIGOScientific:2019vic} and atom interferometers~\cite{Graham:2017pmn,Bertoldi:2019tck}, altogether spanning $\sim 20$ decades in frequency, from $\sim 10^{-17}$ Hz to $\sim$ kHz. Furthermore, a high-frequency detection program at $\sim$ MHz frequencies and above, has just been recently put forward~\cite{Aggarwal:2020olq}. An evaluation of the ability of many of these experiments to probe new high energy physics is still in progress. 

\begin{figure*}[t]
    \begin{center}
    \includegraphics[width=0.31\textwidth]{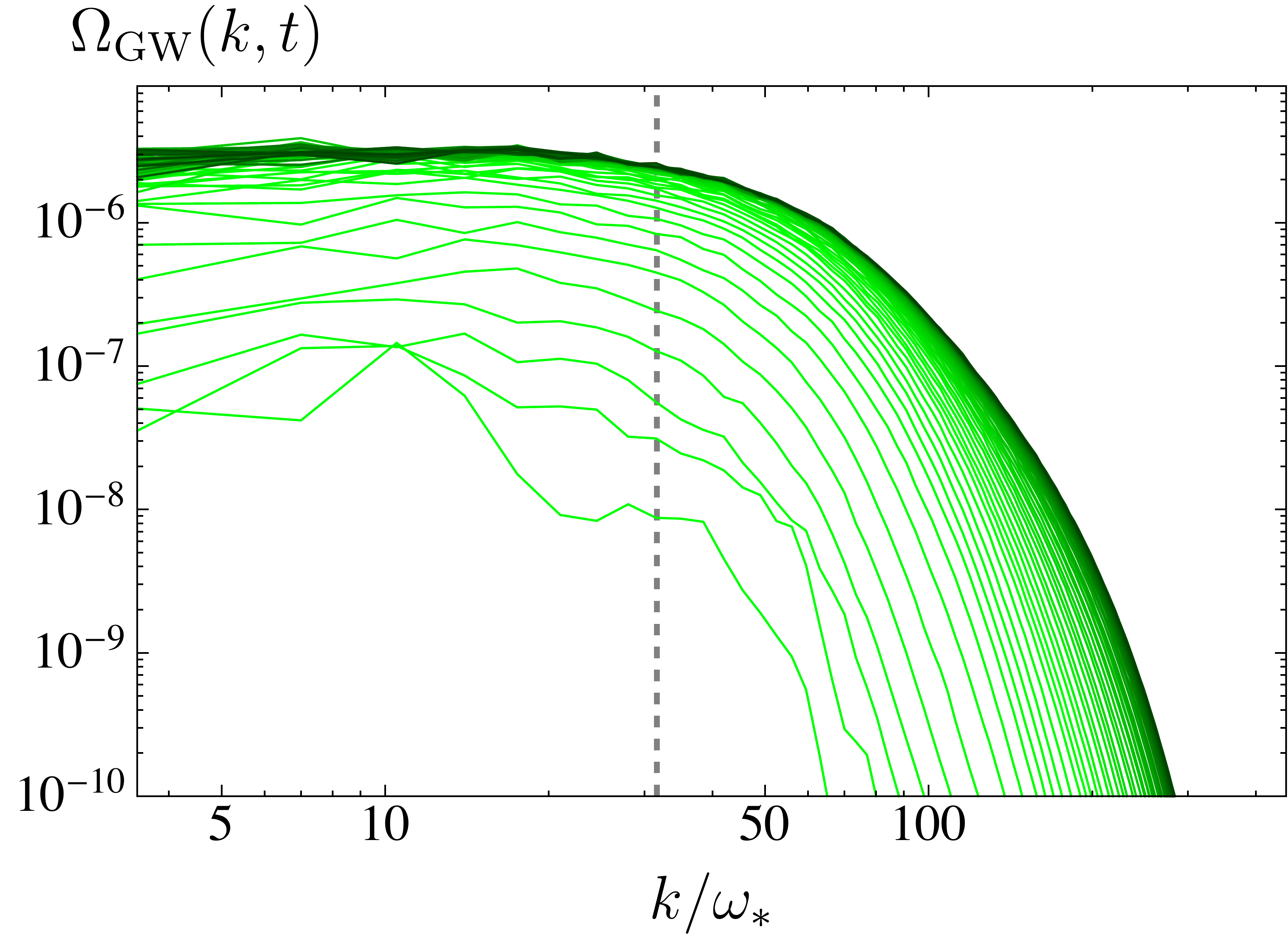} \hspace{0.2cm}
    \includegraphics[width=0.31\textwidth]{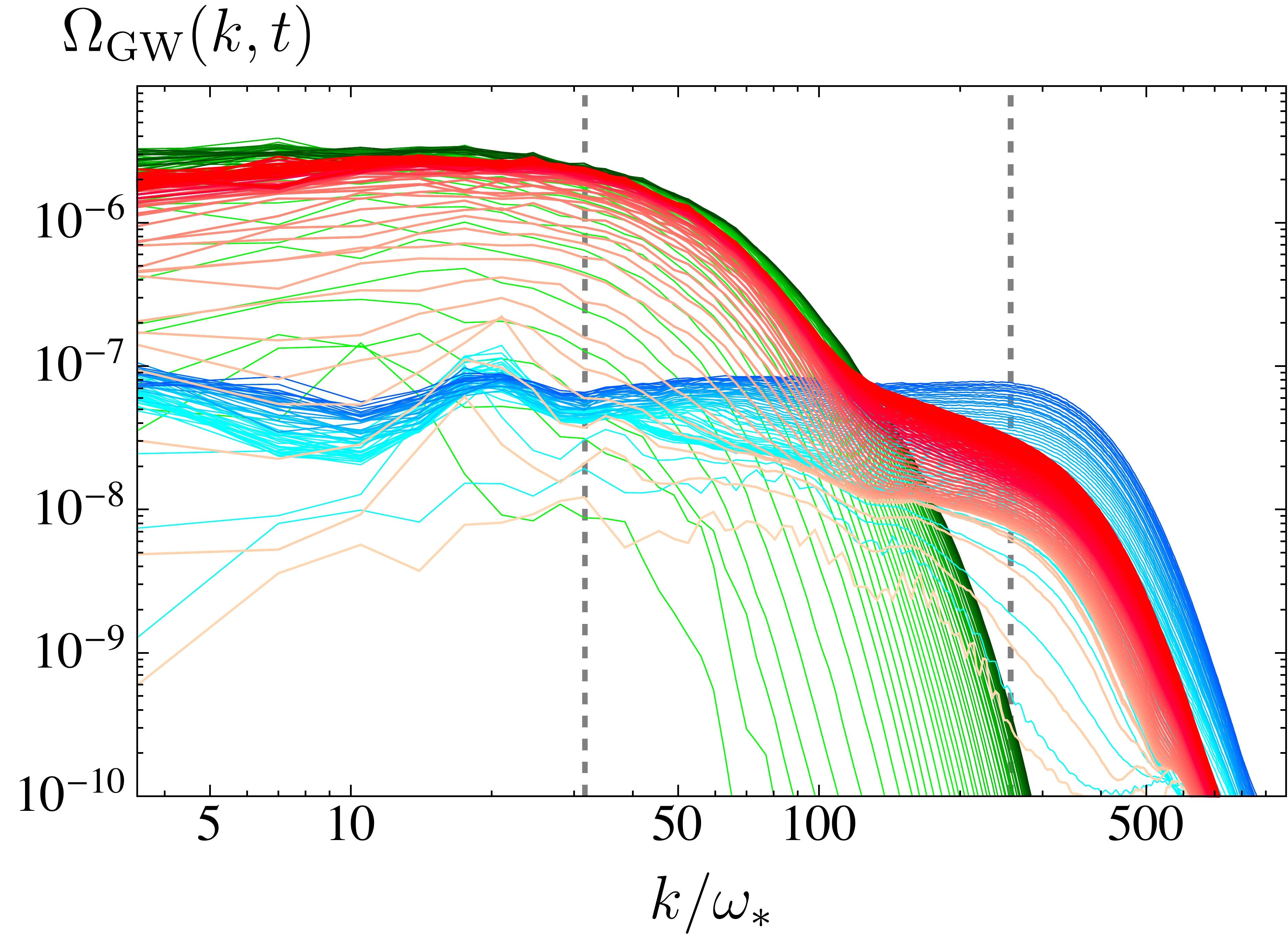}\hspace{0.2cm}
    \includegraphics[width=0.31\textwidth]{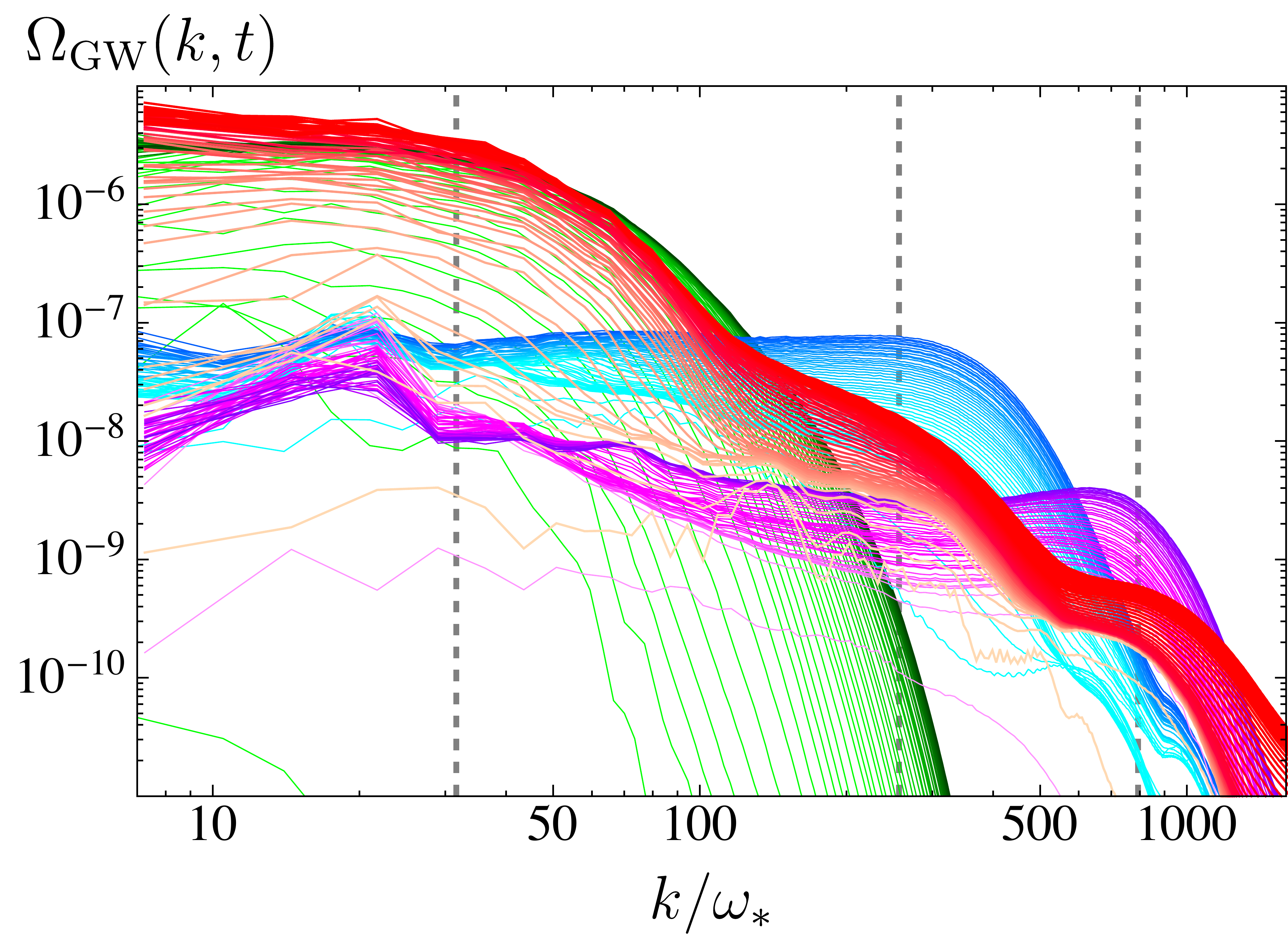} \vspace*{-0.2cm} 
    \end{center}
     \caption{{\it Left}: GW spectrum from a single daughter field with $q = 3\times10^4$. {\it Middle}: GW spectra from single daughter field with $q_1 = 3\times10^4$ (green) and $q_2 = 1.5 \times 10^6$ (blue). A two-peak spectrum (red) emerges when the two daughter fields are simultaneously coupled to the inflaton, with the  same  couplings  as in the  single daughter cases. {\it Right}: GWs from three single daughter fields for $q_1 = 3\times10^4$ (green), $q_2 = 1.5 \times 10^6$ (blue) and $q_3 = 1.3 \times 10^7$ (purple). A three-peak spectrum appears when the three daughter fields are simultaneously coupled to the inflaton, with the same  couplings as the  single daughter cases. In all plots vertical dashed lines show the prediction of Eq.~(\ref{eq:kp(q)law}) for the peak position from single daughter simulations.} \label{fig:Peaks} \vspace*{-0.3cm}
\end{figure*}
In this letter we propose to measure particle couplings with GWs. While in particle colliders one looks for features in statistical observables, we look instead for features in SGWB spectra. Considering early Universe scenarios sourcing GWs, we address the question of how the presence of multiple fields, all contributing to GW production, affect the SGWB spectrum. This question does not have a unique or immediate answer. Indeed, while the source of GWs is a linear superposition of individual contributions, the SGWB spectrum is quadratic in the source. Furthermore, the dynamics of the different fields in the multi-field system may differ from their single-field dynamics. Under certain circumstances, however, a sequence of peaks of different amplitude and frequency -- a {\it stairway} --, may emerge in the SGWB spectrum, probing the leading interaction of each individual species. Detecting such signature would allow to reconstruct the relevant couplings (spectroscopy) of the field species involved in high energy phenomena generating SGWBs.

As a {\it proof of principle}, we consider a preheating scenario where an inflaton $\phi$ oscillates around the minimum of its potential, while multiple {\it daughter} fields $\lbrace \varphi_j \rbrace$ are coupled to the inflaton, each with different coupling strength. When the couplings are large enough, the daughter species are excited in broad parametric resonance, developing spectra that are peaked at scales determined by their coupling strength. By studying the SGWB sourced by various daughter fields, we show that, 
in fact, a stairway of peaks of different amplitude and frequency, emerges in the SGWB spectrum. As each peak probes a different coupling, the detection of such 
signature allows for a reconstruction of the coupling constants. Our working scenario represents a compelling case of the ability of SGWBs to probe particle couplings.\vspace*{0.1cm}\\
{\bf Stairway in the SGWB}. We assume summation over repeated indices and consider a spatially flat Friedmann--Lema\^itre--Robertson--Walker (FLRW) background with tensor perturbations, $ds^2 = -dt^2 + a^2(t)(\delta_{ij} + h_{ij})dx^idx^j$, with $a(t)$ the scale factor, and the perturbations verifying $h_{ii} = \partial_ih_{ij} = 0$. We denote as $m_p = 1/\sqrt{8\pi G} = 2.44\cdot 10^{18}$ GeV the reduced Planck mass. 

\begin{figure*}[t] \begin{center}
    \includegraphics[width=0.28\textwidth,height=4.5cm]{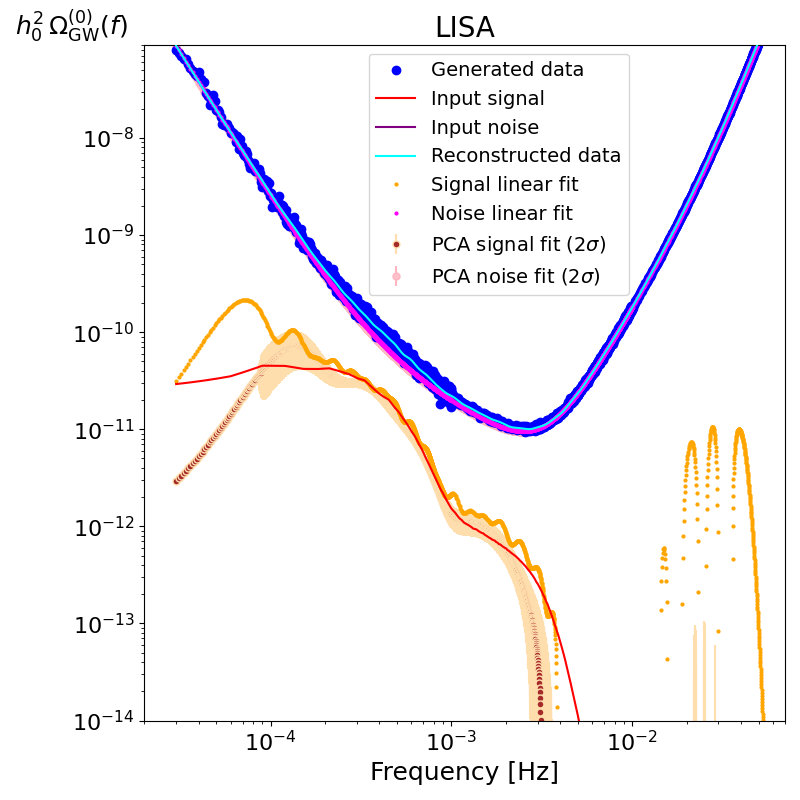} \hspace{0.5cm}
    \includegraphics[width=0.28\textwidth,height=4.5cm]{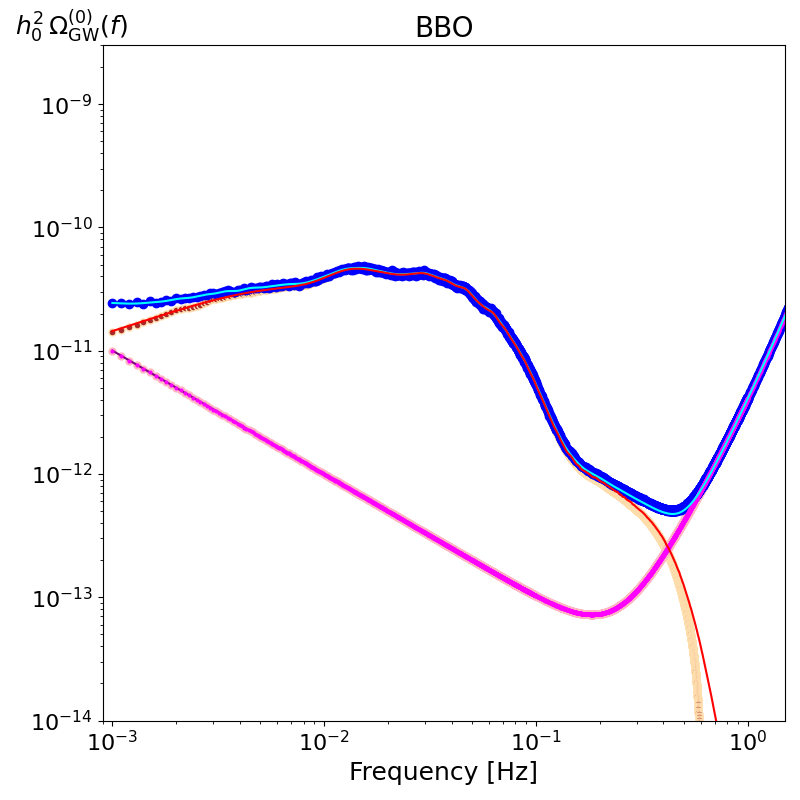} \hspace{0.5cm}
    \includegraphics[width=0.28\textwidth,height=4.5cm]{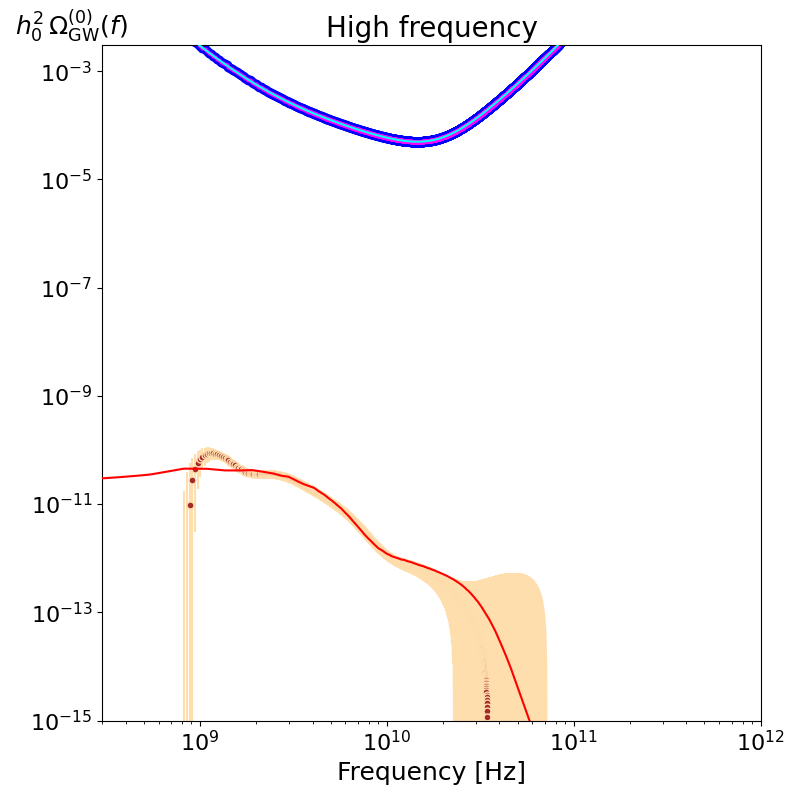} 
    \end{center}
    \vspace*{-0.5cm}
    \caption{Signal and noise reconstructions using Principal Component Analysis over a two-peak GW signal with $q_1 = 3\cdot 10^4$ and $q_2 = 1.5\cdot 10^6$, for LISA (left panel), BBO (central panel) and a speculative High Frequency Experiment (HFE, right panel).} \label{fig:PCA_plots} \vspace*{-0.3cm}
\end{figure*}

We consider an observationally viable inflationary model inspired by $\alpha$-{\it attractors} \cite{Kallosh:2013hoa},
with the inflaton $\phi$ coupled to a set of daughter scalar fields $\lbrace \chi_j \rbrace$,
\begin{eqnarray}\label{eqn:Potential}
   V(\phi,\lbrace \chi_j \rbrace) = \dfrac{1}{2}\Lambda^4 \tanh^2{\left(\dfrac{\phi}{M}\right)} + \dfrac{1}{2}g_j^2 \chi_j^2\phi^2 \,.
\end{eqnarray}
We fix $M = 5 \,m_p$ and $\Lambda = 0.00564 \,m_p$ for compatibility with CMB constraints, and keep free the coupling constants $g_j^2$. The inflaton potential flattens out for $|\phi| \gg M$, and takes the form $V \simeq {1\over2}\omega_*^2 \phi^2$ for $|\phi| \ll M$, with $\omega_* = \Lambda^2 / M$. We choose quadratic interactions as these are scale-free and serve as a proxy for the leading term in scalar-gauge interactions~\cite{Figueroa:2015rqa}. Preheating takes place via \textit{broad parametric resonance}, as the oscillations of the homogeneous inflaton induce an exponential growth of the preheat mode functions ${\chi}_j(k,t)$. The resonance of each species is controlled by their resonance parameter $q_j = g_j^2 \phi_{*}^2 / \omega_*^2 \gg 1$, with $\phi_* \simeq 0.95 \,m_p$ the inflaton amplitude at the end of inflation. Modes of each preheat species are exponentially excited during the initial linear regime, up to a co-moving scale $k \lesssim k_l^{(j)} \sim \omega_{*}q_j^{1/4}$~\cite{Kofman:1997yn}. Eventually, the preheat fields back-react into the inflaton, and the dynamics become non-linear. As a result, mode-to-mode interactions populate higher momenta beyond the linear threshold, typically up to a peak scale $k_{p}^{(j)} \sim \omega_*q_j^p$, $p\gtrsim0.5$~\cite{Figueroa:2017vfa}. Modes with $k > k_{p}^{(j)}$ are exponentially suppressed. The system reaches eventually a stationary state characterised by no significant transfer of energy among species~\cite{Figueroa:2016wxr,Antusch:2020iyq,Antusch:2021aiw}. 

The dynamics of the fields lead to an anisotropic stress with non-vanishing transverse-traceless (TT) part $\Pi_{ij}^{\rm TT} = \{\partial_i \phi \partial_j \phi\ + \partial_i \chi_a \partial_j \chi_a\}^{\rm TT}$. 
GWs are then sourced by all field species, 
with the GW dynamics governed by $(\partial^2_t+3H\partial_t - a^{-2}\nabla^2) h_{ij} = 2m_p^{-2}a^{-2}\Pi_{ij}^{\rm TT}$, with $H = \dot a/a$ the Hubble rate. The SGWB energy density spectrum, normalized to the critical energy density $\rho_c$, is given by
\begin{eqnarray}\label{eqn:GWpowerspectrum}
\Omega_{\rm GW}(k,t) \equiv \dfrac{1}{\rho_c}\frac{d\rho_{\rm GW}}{d\log k}\,,~~~ \frac{d\rho_{\rm GW}}{d\log k} =  \frac{k^3\mathcal{P}_{\dot h}(k,t)}{(4\pi)^3G}\,,\\
\langle\dot{{h}}_{ij}\left(\mathbf{k},t\right)\dot{{h}}_{ij}^{*}\left(\mathbf{k'},t\right)\rangle \equiv (2\pi)^3\,\mathcal{P}_{\dot h}(k,t)\,\delta({\bf k}-{\bf k}')\,,
\end{eqnarray}
with $\langle ... \rangle$ representing stochastic averaging~\cite{LatticeSpectrum}. In this scenario GWs are actively sourced untill the start of the stationary regime, and propagate freely afterwards.  

The non-linear dynamics of the system can only be studied with lattice simulations. We have added for this purpose GW dynamics into the package {\tt ${\mathcal C}$osmo${\mathcal L}$attice}~\cite{Figueroa:2021yhd,moduleGWs}. By running simulations with a single daughter field for different coupling constants $g^2$, we have parametrized the peak's position ($k_p$) and amplitude ($\Omega_{\rm GW}^{(p)} \equiv \Omega_{\rm GW}(k_p)$) of single-daughter SGWB spectra as a function of $q = {g^2\phi_*^2/\omega_*^2}$. By measuring spectra at a time $\omega_{*}t_{\rm f} = 600$, we guarantee (for all couplings used) that the system is in the stationary regime when GW production ceases. We obtain (for $q \gtrsim 10^4$)
\begin{eqnarray}\label{eq:kp(q)law}
k_p(q)/\omega_* = (21.22 \pm 7.64)\left(\dfrac{q}{10^4}\right)^{0.52\pm0.08}~~~~~\,,\\
\label{eq:OmegaGWs(q)law}
\Omega_{\rm GW}^{(p)}(q) = (1.2 \pm 0.7)\times 10^{-5} \left(\dfrac{q}{10^4}\right)^{-1.1\pm0.12}\,,
\end{eqnarray}
It is clear from Eqs.~(\ref{eq:kp(q)law}), (\ref{eq:OmegaGWs(q)law}) that sufficiently separated values of $q$ lead to single-daughter spectra with quite different amplitudes and peak frequencies. In realistic scenarios we naturally expect various daughter fields to be present. If these are coupled to the inflaton with different coupling strengths, relevant questions can then be raised: will the final SGWB spectrum be the superposition of each individual daughter's spectrum? do interference effects affect the final spectral shape? can we expect a series of peaks to emerge at the positions indicated by single daughter field spectra? 

In Fig.~\ref{fig:Peaks} we plot SGWB spectra from single daughter simulations (left, central and right panels), and superimpose the spectrum from simulations with two (central panel) or three (right panel) daughter fields, with the same couplings as the single daughter cases. The plots are self-explanatory: a {\it stairway} signature consisting of a series of peaks, as many as the daughter fields, emerges clearly in the multi-daughter spectra. As the energy must be split between the different species, the stairway peak amplitudes are smaller than in the individual daughter field spectra. The position of the stairway peaks, however, is still well predicted by the single-daughter spectra prediction Eq.~(\ref{eq:kp(q)law}). Redshifting the background, the peak frequencies of a signal today is approximated as~\cite{FreqRedshift}
\begin{eqnarray}\label{eq:fp(q)lawToday}
f_{\rm p} \simeq  4\cdot 10^{10} \times 
\left({H_{\rm f}\over m_p}\right)^{\hspace{-1mm}1/2}\hspace{-2mm}\times {k_p(q)\over a_{\rm f}H_{\rm f}}\,\,~{\rm Hz}\,.
\end{eqnarray}
where $H_{\rm f} = H(t_{\rm f})$ and $a_{\rm f} = a(t_{\rm f})$. To separate clearly the peaks we need $q_{j}/q_{i} \gtrsim {\rm few} \times 10$. It is also important that interactions among daughter species are negligible during GW production (in our calculations we  simply neglected daughter-to-daughter interactions).

\vspace*{0.1cm}

\noindent {\bf Particle coupling spectroscopic reconstruction}.
Using data analysis techniques~\cite{Caprini:2019pxz,Pieroni:2020rob,Flauger:2020qyi} we reconstruct next multi-peak SGWB spectra generated with our simulations, and turn the reconstruction into constrains on the particle couplings thanks to Eqs.~(\ref{eq:kp(q)law}), (\ref{eq:fp(q)lawToday}). 
\begin{figure*}
    \includegraphics[width=0.4\textwidth,height=7.0cm]{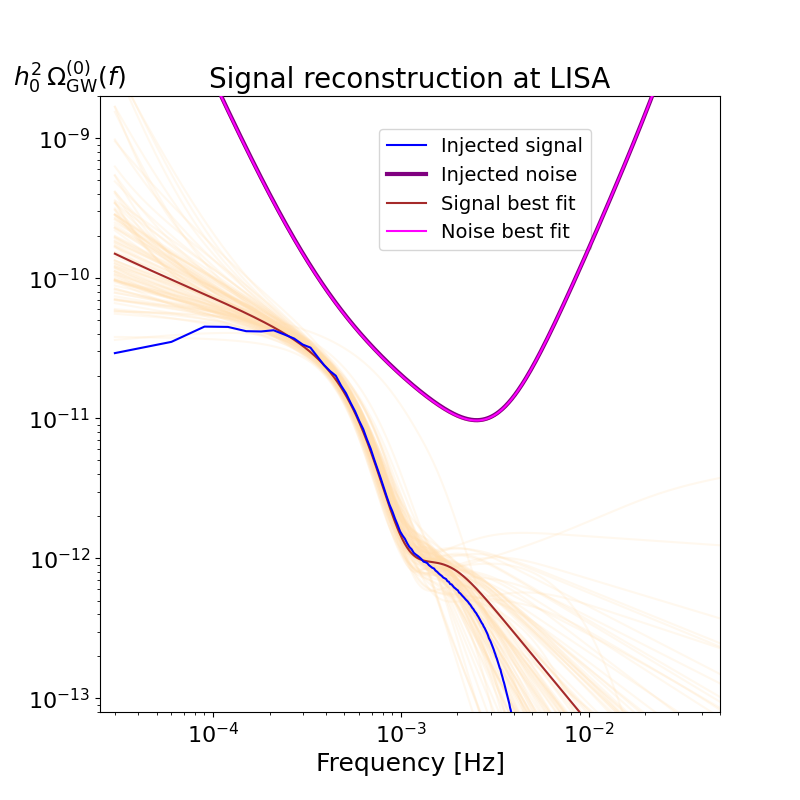} \hspace{0.5cm}
    \includegraphics[width=0.4\textwidth,height=6.5cm]{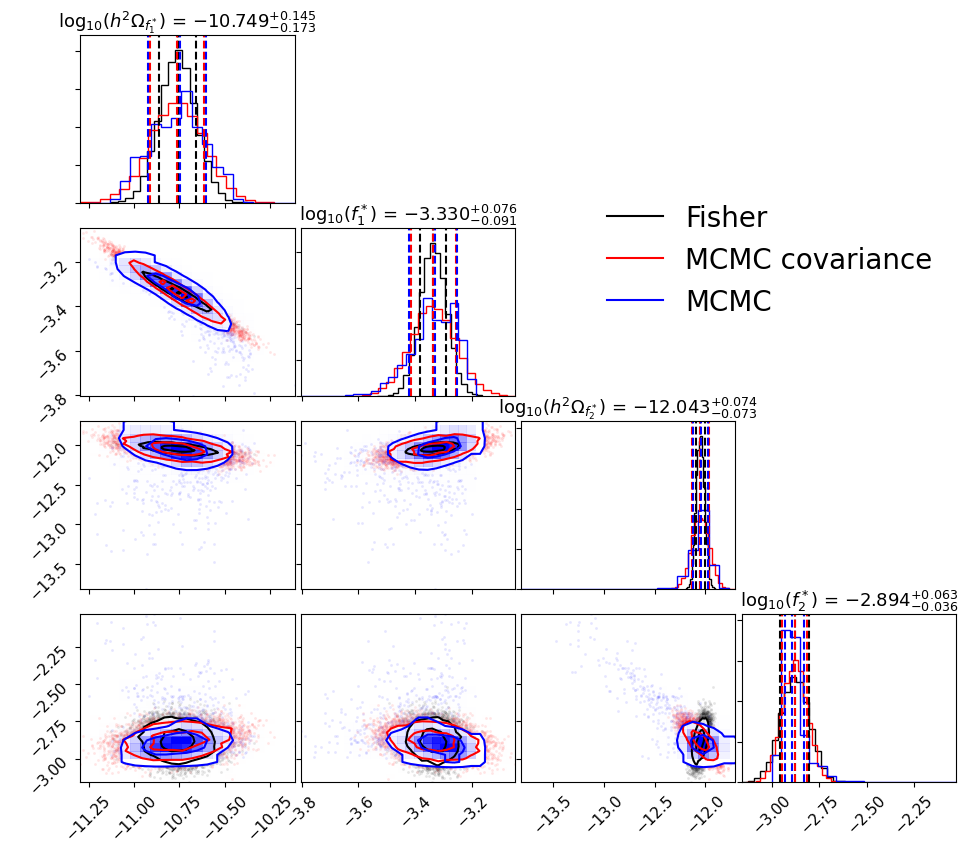} 
    \caption{{\it Left}: Signal and noise reconstruction of a double peaked signal at LISA (see main text for details). The $2\sigma$ signal and noise error bands are shown in orange and pink respectively. {\it Right}: Corner plot for some of the signal parameters showing the $1$ and $2\sigma$ regions. The blue curves are the MCMC chains results, the red curves correspond to the covariance matrix obtained from the MCMC chains and in black we show the constrains obtained with a Fisher matrix approach.} \label{fig:MCMC_fisher} 
\end{figure*}
After generating data sets from our GW signals (see supplementary material), we first apply the Principal Component Analysis (PCA) method of~\cite{Pieroni:2020rob}. This is a model independent procedure for reconstructing an unknown signal by first expanding it onto a general basis of functions, and then dropping the low information components of the Fisher Information Matrix (FIM), $F_{\alpha \beta} \equiv \langle - \frac{\partial^2 \ln \mathcal{L}}{\partial \theta^\alpha \partial \theta^\beta } \rangle$.

Plugging Eq.~(\ref{eq:kp(q)law}) into Eq.~(\ref{eq:fp(q)lawToday}), leads to SGWB peak frequencies around $\sim 100$ MHz for $q_j \sim 10^4-10^6$. These frequencies are way above the range accessible to present GW direct detection experiments like LIGO~\cite{LIGOScientific:2014pky}, Virgo~\cite{VIRGO:2014yos} and KAGRA~\cite{Kawamura:2011zz}, planned  detectors  like LISA~\cite{LISA:2017pwj}, ET~\cite{Punturo:2010zz}, or even futuristic projects like BBO~\cite{Harry:2006fi}. While other multi-field scenarios might produce SGWBs with stairway signatures at observable frequencies, like hybrid preheating scenarios~\cite{Dufaux:2010cf,Tranberg:2017lrx,Cui:2021are} or first order phase transitions~\cite{Kamionkowski:1993fg,Caprini:2007xq,Huber:2008hg,Hindmarsh:2013xza,Hindmarsh:2015qta,Caprini:2015zlo,Hindmarsh:2017gnf,Cutting:2018tjt,Cutting:2018tjt,Cutting:2019zws,Pol:2019yex,Cutting:2020nla} with gauge fields, any characterization of their potential spectroscopic signatures requires dedicated studies currently unavailable. In order to proceed with a coupling reconstruction, we consider instead, for simplicity, that our multi-peak SGWB spectra serve as a template for lower-frequency signal scenarios. We thus shift by hand today's SGWB spectra by a {\it fudge} factor $\alpha_{\rm exp} \equiv f_{\rm exp}/f_{p}$, while leaving their amplitudes intact. The shifted backgrounds peak therefore at some typical frequency $f_{\rm exp}$ of a given experiment. By choosing $\alpha_{\rm LISA} = 1.11 \cdot 10^{-13}$ and $\alpha_{\rm BBO} = 1.55 \cdot 10^{-11}$, we analyze the capability of LISA and BBO to reconstruct SGWB spectra with stairway structures given by our simulation-generated templates. Furthermore, we also introduce a speculative futuristic high-frequency experiment (HFE) capable of detecting our SGWBs at their natural high-frequencies.

Examples of a 2-peak signal reconstruction are shown in Fig~\ref{fig:PCA_plots}, performed respectively with: {\it i)} LISA, using the noise model from~\cite{Flauger:2020qyi, Babak:2021mhe}, {\it ii)} BBO, using the noise model from~\cite{Crowder:2005nr, Harry:2006fi}, and {\it iii)} a futuristic high frequency experiment (HFE), the sensitivity of which we construct {\it ad hoc} in such a way that the signal has roughly the same signal-to-noise ratio as in LISA when shifting the LISA noise curve, see Fig~\ref{fig:PCA_plots}. The two peaks in the signal are clearly identified in the three cases. By using our procedure, together
with alternative methods~\cite{Caprini:2019pxz, Flauger:2020qyi}, it is possible to reconstruct the SGWB frequency shape in an agnostic way. Once an appropriate shape has been identified, a direct search can be actually used in order to directly constrain the model parameters. For this purpose we use the same likelihood as in Eq.(3.12) of~\cite{Flauger:2020qyi}, and model a general $n$-peaked signal as a sum of power law signals with exponential cut-offs:
\begin{equation}
\label{eq:GW_template}
    h^2 \Omega_{GW} = \sum_{i = 1}^n 10^{\ln\left( h^2 \Omega^*_{i}\right)} \frac{ 2 \left( \frac{f}{f^*_{i}} \right)^{n_{i}}  }{ 
                1 + \exp \left\{ \frac{\delta_i (f - f^*_{i})}{f^*_{i}} \right\} } \; , 
\end{equation}
where $h_0 \equiv H_0$[Mpc/Km/s]$/100$ characterises today's Hubble rate, $h_0^2 \Omega^*_{i}$ denotes the amplitude of the $n$-th signal in the peak frequency $f^*_{i}$, $n_{i}$ its power-law tilt, and $\delta_{i} > 0$ controls the strength of the exponential suppression. In Fig.~\ref{fig:MCMC_fisher} we show the results obtained by injecting the same signal used in the left panel of Fig~\ref{fig:PCA_plots}, and simultaneously fitting for signal and noise with LISA. The parameter space is sampled using~\texttt{Polychord}~\cite{Handley:2015fda, Handley:2015vkr} via its interface with~\texttt{Cobaya}~\cite{Torrado:2020dgo}. In the left panel we show the reconstruction of the signal and of the noise as compared to the injections, and in the right panel we show the contour plots for the amplitudes and pivot frequencies of the two-peak spectra (after marginalizing over all other parameters). For reference we compare with results obtained with a Fisher matrix approach which only includes the signal parameters. As seen in the figure, the deviation between the two results is minimal, what implies that the inclusion of noise parameters does not significantly affect the reconstruction of the signal. 

In the three cases we reconstruct the coupling constants as indicated in Table~\ref{tab:1}, for a GW signal obtained from a two-daughter field simulation with  $q_1 = 3\cdot 10^4$ and $q_2 = 1.5\cdot 10^6$. The determination of the peak frequencies and their errors for each experiment (for BBO and HFE we use a FIM approach), leads to a reconstruction of the resonance parameters inverting Eqs.~(\ref{eq:kp(q)law}) \& (\ref{eq:fp(q)lawToday}), from where we obtain the coupling constants. While the numerical reconstruction of $q_i$'s and $g_i$'s is only precise to within the order of magnitude, it is remarkable that all reconstructed values are compatible at $1\sigma$ with the injected values. This clearly indicates the great potential for particle coupling spectroscopy that SGWB detection may offer in the future.

\begin{table}[t]
\hspace*{-0.2cm}{\footnotesize
\begin{tabular}{|c | c | c|  c |} 
\hline
 {} & {\bf LISA} & {\bf BBO} &  {\bf HFE} \\
 \hline
 {} & {} & {} & {}\\[-2.5ex]\hline \\[-2ex]
 $\log_{10}(f_1^*{\rm [Hz]})$ & $-3.33^{+0.08}_{-0.09}$ & $-1.539^{+0.002}_{-0.002}$ & $9.45^{+0.15}_{-0.15}$  \\ [1ex]
 {} & {} & {} & {}\\[-2.5ex]\hline \\[-2ex]
$ q_1 = 3\cdot10^4 $& $6.14^{+29.86}_{-4.10} \cdot 10^{4}$ & $1.28^{+2.45}_{-0.62} \cdot 10^{4}$  & $2.83^{+17.88}_{-2.06} \cdot 10^{4}$ \\ [1ex]
 {} & {} & {} & {}\\[-2.5ex]\hline \\[-2ex]
 $g_1 = 1.16\cdot10^{-3}$ & $1.66^{+4.01}_{-0.55} \cdot 10^{-3}$ & $0.76^{+0.73}_{-0.18} \cdot 10^{-3}$  & $1.13^{+3.56}_{-0.41} \cdot 10^{-3}$ \\ [1ex]
 {} & {} & {} & {}\\[-2.5ex]\hline \\[-2ex]
 $\log_{10}(f_2^*{\rm [Hz]})$ & $-2.89^{+0.06}_{-0.04}$ & $-0.496^{+0.001}_{-0.001}$  & $10.4^{+0.4}_{-0.4}$ \\ [1ex]
 {} & {} & {} & {}\\[-2.5ex]\hline \\[-2ex]
 $q_2 = 1.5\cdot10^{6}$ & $0.43^{+2.8}_{-0.3} \cdot 10^{6}$  & $1.3^{+7.42}_{-0.89} \cdot 10^{6}$ & $1.9^{+108.65}_{-1.78} \cdot 10^{6}$ \\ [1ex]
 {} & {} & {} & {}\\[-2.5ex]\hline \\[-2ex]
 $g_2 = 8.2\cdot10^{-3}$ & $4.39^{+14.3}_{-1.53} \cdot 10^{-3}$ & $7.64^{+21.8}_{-2.61} \cdot 10^{-3}$ & $9.23^{+263.9}_{-4.32} \cdot 10^{-3}$ \\ [1ex]\hline
\end{tabular}
\caption{Numerical reconstruction of frequency peaks, resonance parameters and coupling constants.}
\label{tab:1}
}
\end{table}

{\bf Discussion.--} Realistic early universe scenarios are naturally expected to contain multiple fields, as there is no symmetry or fundamental principle that implies otherwise. The Standard Model (SM), between matter and gauge fields, contains dozen of fields. Beyond the SM (BSM) scenarios only add up to this argument. In this Letter, we discuss the possibility to measure particle couplings through SGWBs. We focus on the emergence of a sequence of peaks of different amplitude and frequency, a {\it stairway}, in the SGWB spectrum produced in scenarios with  multiple particle species involved in the GW generation. As a proof of principle we have studied a preheating scenario with an oscillating inflaton $\phi$, coupled simultaneously to multiple {\it daughter} fields $\lbrace \varphi_j \rbrace$, interacting though $g_j^2\varphi_j^2\phi^2$ with different coupling strengths. When the daughter species are excited, we show that a stairway signature with as many peaks as fields $\lbrace \varphi_j \rbrace$, is imprinted in the SGWB spectrum, with each peak probing the interaction strength of each individual species. 

Our working scenario represents a compelling case of the ability of SGWBs to probe particle couplings. We have assumed a quadratic potential $V \propto \phi^2$ during the inflation oscillations, so that only daughter field peaks are imprinted. If a different shape was considered, say e.g.~$V \propto \phi^p$ with $p \neq 2$ or a linear combination of monomials, inflaton self-resonance will possibly create new peaks in the SGWB spectrum. For simplicity we have assumed the same type of quadratic interactions $\propto \varphi_j^2\phi^2$ for all daughter species. 
We expect however the emergence of analogous stairway signatures in the SGWB spectrum when other interactions are considered, like those in tachyonic preheating \cite{Felder:2000hj,Felder:2001kt,GarciaBellido:2002aj,Copeland:2002ku,Dufaux:2010cf,Tranberg:2017lrx}, axion preheating~\cite{Adshead:2015pva,Cuissa:2018oiw,Adshead:2019lbr,Cui:2021are}, or geometric preheating~\cite{Bassett:1997az,Tsujikawa:1999jh,Fu:2019qqe,Figueroa:2021iwm}. In the case of strong first order phase transitions (see e.g.~\cite{Hindmarsh:2020hop} for a review), gauge fields coupled to the corresponding Higgs sector of the theory will be inevitably excited during the nucleation and collision of bubbles. Arguably this should lead to new peaks in the SGWB spectrum probing the Higgs-gauge couplings. 

We note that in our numerical calculations we have assumed that interactions among the daughter species are negligible within the time scale of GW generation. If daughter-to-daughter interactions were active during GW production, this would affect their dynamics, for example re-distributing power differently among them. The presence of such interactions may therefore affect significantly the peak structure of the resulting stairway in the SGWB spectrum, an aspect we hope we to return to in a future investigation.

{\it Acknowledgements--.} We thank Robert Plant, Jimmy Page, John Paul Jones and John Bonham for suggesting a nice title for our paper and for providing great inspiration and good times. DGF is supported by a Ram\'on y Cajal contract with Ref.~RYC-2017-23493. AF is supported by the U.S. Department of Energy, Office of Science, Office of Nuclear Physics, grants Nos. DE-FG88ER40388. NL is supported by project PROMETEO/2019/083 from Generalitat Valenciana. The work of M.P. was supported by STFC grants ST/P000762/1 and ST/T000791/1. M.P. acknowledges support by the European Union’s Horizon 2020 Research Council grant 724659 MassiveCosmo ERC- 2016-COG. This work is also supported by PROMETEO/2021/083 from Generalitat Valenciana, and by PID2020-113644GB-I00 from Ministerio de Ciencia e Innovaci\'on. We are also very thankful for the use of computational resources provided by Finis-Terrae II cluster of CESGA (Centro de Supercomputación de Galicia), Vives and Tirant clusters of Universitat de Valencia.

\section{Supplemental material}
{\bf Background redshift}. The SGWB today is obtained by redshifting the amplitude and frequency of the background from the end of GWs production $t_{end}$,
following the appropriate expansion history of the Universe. Given that we know the expansion history up to the final time $t_{\rm f}$ of our simulations, and noticing that $t_{end} < t_{\rm f}$, we can apply the redshifting from $t_{\rm f}$ until the present time $t_0$. Let us characterize the expansion history between $t_{\rm f}$ and the onset of radiation domination (RD) at $t_{\rm RD}$, with an effective equation of state $\bar{w}=p/\rho$
\begin{equation}
\begin{array}{cc}
    \log\left(\dfrac{\rho^{\rm (RD)}_{tot}}{\rho^{\rm (f)}_{tot}}\right) & = -3 \int^{a_{\rm RD}}_{a_{\rm f}} \dfrac{da'}{a'}(1+w(a')) \: ,  \\
     & = -3(1+\bar{w}) \log\left(\dfrac{a_{\rm RD}}{a_{\rm f}}\right) \: ,
\end{array}
\end{equation}
The SGWB spectrum actually peaks at some sub-horizon scale $k_p = \beta_p a_{\rm f} H_{\rm f} $, with $a_{\rm f} = a(t_{\rm f})$ and $H_{\rm f }= H(t_{\rm f})$. We expect the SGWB spectrum to peak therefore at a frequency $f_{\rm GW}$ today
\begin{eqnarray}
f_{\rm GW} &\equiv& \dfrac{k_{\rm GW}}{2\pi a_0} = \dfrac{\beta_{\rm GW}}{2\pi} \left(\dfrac{a_{\rm f}}{a_{\rm RD}}\right)\left(\dfrac{a_{\rm RD}}{a_0}\right) H_{\rm f} \nonumber\\
     &=& \dfrac{\beta_{\rm GW}}{2\pi} \left(\dfrac{a_{\rm f}}{a_{\rm RD}}\right)G_{\rm RD}^{1/4}\left(\dfrac{\rho_{rad,0}}{\rho_{\rm rad,RD}}\right)^{1/4} H_{\rm f}  \nonumber\\
     &=& \dfrac{\beta_{\rm GW}}{2\pi} (2 \epsilon_{\rm f})^{1/4}G_{\rm RD}^{1/4}\left(\dfrac{H_{\rm f}}{\rho_{\rm rad,RD}}\right)^{1/4} \rho_{\rm rad,0}  \nonumber\\
     &=& \dfrac{k_p}{a_{\rm f}H_{\rm f}}  \epsilon_{\rm f}^{1/4}G_{\rm RD}^{1/4}\left(\dfrac{H_{\rm f}}{m_p}\right)^{1/2} \left(\dfrac{2}{3}\right)^{1/4}  \dfrac{\rho_{\rm rad,0}^{1/4}}{2 \pi} \, ,
\end{eqnarray}
where we have defined 
\begin{equation}
\hspace*{-0.4cm}\epsilon_{\rm f} \equiv \left(\dfrac{a_{RD}}{a_{\rm f}}\right)^{1-3\bar{w}}\,,~~ G_{RD} \equiv \left(\dfrac{g_{RD}}{g_{0}}\right)\left(\dfrac{g_{s,0}}{g_{s,RD}}\right)^{4/3}\hspace*{-0.1cm},
\end{equation}
with $g_{s,t}$ and $g_{t}$ the entropic and energy density relativistic degrees of freedom. We can characterize the factor $G_{\rm RD}^{1/4}$ taking into account that the SM degrees of freedom above the electro-weak scale amount to $g_{s,RD}=g_{RD}=106.75$, so we write $G_{\rm RD}^{1/4}\simeq (g_{s,RD}/100)^{-1/12}$. The redshift factor finally reads, using $\rho_{rad,0} = 3.37\times10^{-51}\text{GeV}^4$, 
\begin{equation}
    f_{\rm GW} \simeq 4\times 10^{10} \, \epsilon_{\rm f}^{1\over4} \left(\dfrac{g_{\rm s,RD}}{100}\right)^{-{1\over12}} \: \dfrac{k}{a_{\rm f} H_{\rm f}} \left(\dfrac{H_{\rm f}}{m_p}\right)^{1\over2} \: \text{Hz} \: .
\end{equation}
The redshifted GW spectrum amplitude is 
\begin{equation}
        h_0^2\Omega_{\rm GW}^{(0)}(f) = \epsilon_{\rm f} \:G_{\rm RD} \:h_{0}^2 \Omega_{rad}^{(0)} \:\Omega_{\rm GW}(k) \:, 
\end{equation}
and using the value of $h_{0}^2 \Omega_{\rm rad}^{(0)} \approx 4\times10^{-5}$, the peak amplitude of the spectrum today is
\begin{equation}
        h_0^2\Omega_{\rm GW}^{(0)}\Big|_{\rm peak} \simeq 1.6 \times 10^{-5} \: \epsilon_{\rm f} \: \left(\dfrac{g_{\rm s,RD}}{100}\right)^{-1/3} \: \Omega_{\rm GW}^{\rm (p)} \: . 
\end{equation}
For simplicity we consider $\epsilon_{\rm f}=1$ and $g_{\rm s,RD}=100$ (i.e.~$G_{\rm RD} = 1$) when redshifting the SGWB from our lattice simulations, as this choice is irrelevant for our major purpose of coupling reconstruction.

{\bf GWs in the lattice.} The energy density of a homogeneous and isotropic GW background is given by~\cite{Caprini:2018mtu}
\begin{equation}\label{eqn:GWenergydensity}
    \rho_{\rm GW}(t) = \dfrac{1}{32\pi G}\langle \dot{h}_{ij}({\bf x},t)\dot{h}_{ij}({\bf x},t)\rangle \: ,
\end{equation}
with $\langle...\rangle$ a spatial average over a large enough volume to encompass all relevant wavelengths. In the limit $V \rightarrow \infty$, the spectrum per logarithmic interval in Fourier space is
\begin{equation}\label{eqn:GWspowerspectrumContinous}
    \dfrac{d \rho_{\rm GW}}{d \log k} = \dfrac{k^3}{(4\pi)^3 G V} \int \dfrac{d \Omega_k}{4\pi} \dot{h}_{ij}({\bf k},t) \dot{h}^*_{ij}({\bf k},t) \: ,
\end{equation}
where $d\Omega_k$ represents a solid angle element in {\bf k}-space. The integral average over the spherical shell of radius $|{\bf k}|$ mimics in the lattice the stochastic average of Eq.~\ref{eqn:GWpowerspectrum}, as long as GWs are sourced by a source with random initial fluctuations. Following reference~\cite{Figueroa:2011ye}, we obtain a discretized version of the aforementioned equation  
\begin{eqnarray}
\left(\dfrac{d \rho_{\rm GW}}{d \log k}\right)({\bf \Tilde{n}}) &=&  \dfrac{k^3(\Tilde{\bf n})dx^3}{(4\pi)^3 G  N^3}\left\langle \dot{h}_{ij}(|\tilde{\textbf{n}}|,t) \dot{h}_{ij}^*(|\tilde{\textbf{n}}|,t)\right\rangle_{R(\Tilde{\bf n})}\,,\nonumber\\
\end{eqnarray}
where $\Tilde{\bf n}$ is a lattice site in the {\it Fourier} space, $k = k_{\rm IR} |\tilde{\bf n}|$ is the momentum modulus, $dx$ is the lattice spacing, $N$ is the number of sites per spatial dimension in the lattice, and $\langle...\rangle_{R(\Tilde{\bf n})}$ is an average over a spherical shell of radius $R(|\Tilde{\bf n}|) = k_{\rm IR} |\Tilde{\bf n}|$ and width $\Delta R = k_{\rm IR}$, with an approximate number of modes $4 \pi |\Tilde{\bf n}|^2$ on the shell $R(|\Tilde{\bf n}|) = \left[|\Tilde{\bf n}|-\Delta \tilde{n}/2,|\Tilde{\bf n}|+\Delta \tilde{n}/2\right)$, where $\Delta \tilde{n} = 1$ and $|\Tilde{\bf n}| = 1, 2 \dots [\sqrt{3}N/2]$. 
 
In order to evolve the $h_{ij}$ fields in the lattice, we follow the method introduced in~\cite{Garcia-Bellido:2007fiu}. We evolve a set of fields $u_{ij}$ with the equation of motion
\begin{equation}\label{eqn:u'sEoM}
    \ddot{u}_{ij}+3H\dot{u}_{ij} - \dfrac{\nabla^2}{a^2} u_{ij} = \dfrac{2}{m_p^2 a^2}\{\partial_i \phi_k \partial_j \phi_k\} \:,
\end{equation}
where $k$ runs over all the fields. The physical transvere-traceless part is obtained through
\begin{equation}\label{eqn:TTprojector}
    h_{ij}(k,t)= \Lambda_{ijkl}(\hat{k})\: u_{kl}(k,t) \:,
\end{equation}
where $\Lambda_{ijkl}$ is defined by
\begin{eqnarray}
\label{eqn:ProjectorTT}
    \Lambda_{ijlm}({\bf \hat{k}}) \equiv P_{il}({\bf \hat{k}})  P_{jm}({\bf \hat{k}}) - \dfrac12 P_{ij}({\bf \hat{k}}) P_{lm}({\bf \hat{k}}) \: ,\\
     P_{ij}= \delta_{ij} - \hat{k}_i \hat{k}_j\:,~~  \hat{k}_i = k_i/k \:. ~~~~~~~~~
\end{eqnarray}
The definition of a lattice projector $\Lambda_{ij,lm}^{\rm (L)}$ is analogous as in Eq.~(\ref{eqn:ProjectorTT}), but in terms of a lattice momentum $k_i^{\rm (L)}$ that depends on the choice of spatial-derivative, see~\cite{Figueroa:2011ye} for a discussion. We use the nearest-neighbor derivative of equation (71) in~\cite{Figueroa:2020rrl}, for which the lattice momenta is given by 
\begin{equation}
    k^{(L)}_{i} = 2 \dfrac{\sin (\pi \Tilde{n}_i/N)}{dx} \: .
\end{equation}

{\bf GW data analysis.} Let us define first the signal-to-noise ratio (SNR) as a way to quantify the strength of a signal with respect to a given experiment noise, as
\begin{equation}
    \textrm{SNR} = \sqrt{ T \int_{f_{\textrm{min}}}^{f_{\textrm{max}}} \left( \frac{\Omega_{\rm GW}^{(0)} (f)}{\Omega_{n} (f)} \right)^2} \; ,
\end{equation}
with $T$ the observation time, $f_{\textrm{min}}$ and $f_{\textrm{max}}$ the minimal and maximal frequencies of the experiment, and $\Omega_{\rm GW}^{(0)}(f)$, $\Omega_{n}(f)$ respectively the signal and the noise power spectra in $\Omega$ units. 

We review now the data generation procedure we have implemented. For any given experiment, we assume that a data stream $d_I(t)$, with $I$ running over different data channels, will be provided in time domain. The observation time $T$ can then be divided into $N_d$ segments of duration $T/N_d$. In the Fourier domain we write $\tilde{d}^i_I(f_\textmd{k})$, where $i$ runs over segments, and $\textmd{k}$ over frequencies in the detector's range. We define the frequency resolution within each segment as $\Delta f = N_d/T$. By assuming different frequencies to be uncorrelated and both the signal and noise to be Gaussian distributed with vanishing mean and variance given by their respective spectra, we can generate $N_d$ statistical realizations of the signal and noise. 
We then define a new set of (averaged) data $\bar{D}^\textmd{k}_{IJ} \equiv 
\tilde{d}^i_J(f_\textmd{k}) \tilde{d}^i_I(f_\textmd{k}) / N_d$ which, we down-sample using a coarse graining procedure~\cite{Caprini:2019pxz, Flauger:2020qyi}. After this we are left with a data set $D^k_{IJ}$, where $k$ runs now over a lesser dense set of frequencies, with a set of weights $w^k_{IJ}$, corresponding to the number of points we average over in the coarse graining procedure. The new down-sampled data have the same statistical properties of the $\bar{D}^\textmd{k}_{IJ}$ but are easier to handle numerically~\cite{Caprini:2019pxz, Flauger:2020qyi}. 

We review next the basics of the PCA technique of~\cite{Pieroni:2020rob} (see also~\cite{Fumagalli:2021dtd}). For simplicity we restrict ourselves to a single data channel per experiment~\cite{Channels} and proceed to build a Gaussian likelihood to describe the data:
	\begin{equation}
		- \ln \mathcal{L} \simeq \frac{N}{2} \sum_k w_k \left[   \frac{ D^k - D^{\rm th}(f_k, \vec{\theta}) }{ D^k} \right]^2 \; ,
	\end{equation}
with $D^{\rm th}(f_k, \vec{\theta})$ the theoretical model (containing both signal and noise) to describe the data, depending on a vector $\vec{\theta}$ of parameters. This likelihood is known to be low-biased~\cite{Bond:1998qg, Sievers:2002tq, WMAP:2003pyh, Hamimeche:2008ai}, but for the scope of our analysis this can be safely ignored. Assuming the model to be linear in $ \vec{\theta}$, then $\ln \mathcal{L}$ is quadratic. As a consequence, for any given the model, finding the maximum likelihood estimate $ \vec{\theta}_b$ of the best-fit parameters reduces to solving a linear problem. Following~\cite{Pieroni:2020rob}, a model independent approach to this problem can be defined by expanding the signal onto some basis (in particular we will use the same basis of Gaussians of~\cite{Pieroni:2020rob}) and solving for the $ \vec{\theta}_b$. By computing the eigensystem of the FIM, it is then possible to cut the low information components, which naturally allows to express the signal only in terms of the well-determined combinations.

Finally we comment on the procedure employed to derive Eq.~(\ref{eq:kp(q)law}) and on its impact on the determination of the couplings in Table~\ref{tab:1}. By taking a set of simulations with a single daughter field, we determine the peak position by fitting the corresponding SGWB spectrum (at the end of our simulations) with the template given in Eq.~(\ref{eq:GW_template}). From this procedure we obtain a set of $k_i$'s which we associate to the $q_i$'s used. Eq.~(\ref{eq:kp(q)law}) is then obtained by performing a power law fit for $k_i(q_i)$. Increasing the number of data points (\emph{i.e.} performing more lattice simulations) the goodness of this fit could be improved, leading to an improvement of the determinations of the $q_i$'s and $g_i$'s given in Table~\ref{tab:1}. Notice that the error bars for the $q_i$'s and $g_i$'s are always non-symmetric. This is a consequence of our choice of using $\ln(f_i^* [Hz])$ as a parameter for the reconstruction of the GW spectrum. In fact a symmetric error band in $\ln(f_i^* [Hz])$ (as the one obtained from a Fisher analysis) always lead to a larger band in the $+$ direction as compared to the $-$ one. 

\bibliography{spect_GW_auto,spect_GW_manual}

\begin{thebibliography}{151}
\bibitem [{Cha()}]{Channels}%
  \BibitemOpen
  \href@noop {} {}\bibinfo {howpublished} {In most cases, and in particular for
  all the cases we consider, this would only lead to a sub-optimal estimation
  (the error scales as $1/\sqrt{N_c}$, $N_c$ being the number of data channels)
  of the parameters. Since we are only using this technique in order to
  identify the SGWB frequency shape, we can safely neglect this
  point.}\BibitemShut {Stop}%
\end{thebibliography}

\end{document}